\documentclass[reprint, amsmath,amssymb,showpacs, aps,prd]{revtex4-1}
\usepackage[utf8]{inputenc}
\usepackage{amsmath}
\usepackage{tikz}
\usepackage{amsfonts}
\usepackage{amssymb}
\usepackage{mathrsfs}
\usepackage{latexsym}
\usepackage{footnote}
\usepackage{hyperref}
\usepackage[english]{babel}
\usepackage{cmap}
\usepackage[T1]{fontenc}
\usepackage{textcomp}
\usepackage{setspace}
\usepackage{graphicx}
\usepackage{dcolumn}
\usepackage{bm}
\usepackage{float}
\begin{document}

\preprint{APS/123-QED}

\title{Searching evidences of Neutrino-Nucleus Coherent Scattering  with M\"ossbauer Spectroscopy}

\author{$^1$$^,$$^2$C. Marques}
\author{$^1$G. S. Dias}
\author{$^3$H. S. Chavez}
\author{$^2$S. B. Duarte}
\affiliation{$^1$Department of Physics - IFES - Instituto Federal do Esp\'{i}rito Santo (IFES), \\ Av. Vit\'{o}ria 1729, Jucutuquara, Vit\'{o}ria, ES, Brazil, CEP 29040-780}
\affiliation{$^2$Centro Brasileiro de Pesquisas F\'{i}sicas (CBPF) Rua Dr. Xavier \\ Sigaud 150, Urca, RJ, Brazil, CEP 22290-180} 
\affiliation{$^3$UCL-Faculdade do Centro Leste, Rodovia ES 010, Km 6 \\ BR 101, Serra-ES, Brazil}

\email{celiom@ifes.edu.br}
\email{gilmar@ifes.edu.br}
\email{helderch@hotmail.com}
\email{sbd.cbpf.rj@gmail.com}
\date{\today}
\begin{abstract}
The M\"ossbauer technique is proposed as an alternative experimental procedure to be used in the detection of Coherent Elastic $\nu$-Nucleus Scattering (CENNS). The $Z^{0}$-boson exchange interaction is considered as a perturbation on the nuclear mean-field potential. This causes a change in the valence neutron quantum states in the $^{57}$Fe nucleus of the M\"ossbauer detector, which is a typical isotope used in M\"ossbauer spectroscopy. The transferred energy causes a perturbation at the valence neutron level, modifying the location of the isomeric peak of the M\"ossbauer electromagnetic resonance. We calculate the CENNS isomeric shift correction and show that this quantity is able to be detected with enough precision. Therefore, the difference between the M\"ossbauer isomeric shift in the presence of a reactor neutrino beam and without the neutrinos flux is pointed out as a figure of merit to manifest CENNS. In this work, we show that the CENNS correction of the isomeric shift is of $\approx 10^{-7}$eV, which is greater than the $10^{-10}$eV resolution of the technique.
\end{abstract}


\maketitle


\section{\label{sec:level1}Introduction}

The search for Coherent Elastic $\nu$-Nucleus Scattering (CENNS) is considered an experimentally challenging task as an  introduction of the weak neutral current in the context of the Standard Model (SM) \cite{freedman,giunt,brice}. An observation has been intensively pursued by many experimental collaborations in the last four decades~\cite{drukier,formaggio,giunt,brice,collar} developing a great effort to detect this weak process with largest predicted cross section. 

However, only recently the COHERENT collaboration detected undoubtedly the process~\cite{akimov} using an efficient scintillator detector. The experiment, located at Oak Ridge National Laboratory, uses a Spallation Neutron Source with an extremely intense neutron beam. In the scattering of these neutrons by a mercury target, a secondary pion beam is produced. The pion decay generates the intense neutrino flux ($\approx 10^{11}/$s) used in the experiment, with energy in the range of $16$ to $53$~MeV~\cite{akimov}. The experiment accumulates fifteen months of photoelectrons, which was produced in accordance to the SM prediction, when the pulsed neutrino flux is scattered by $14.6$ kg crystal made of CsI doped with Sodium atoms. The experimental setup was properly structured to prevent any contamination from external sources of neutrons and neutrinos, like atmospheric or solar and galactic neutrinos.

The present work proposes the study of the CENNS using M\"ossbauer nuclear spectroscopy(MS) \cite{greenwood,gruverman}. Note that the main characteristic of M\"ossbauer technique is the recoilless interaction of the electromagnetic radiation with the nucleus. 

We consider that the nucleus has a fixed position in the local minimum of the crystal potential lattice and that the effect of the transferred energy is to induce a perturbation in the quantum state of the valence neutron in the atomic nuclei. The volume of the nucleus is modified, with the change in electronic density of K-electron inside the nucleus. The magnitude of this change can be measured by the isomeric displacement in a typical M\"ossbauer experiment. We assume that only the valence neutron in $^{57}$Fe  is excited with the $Z^{0}$ nuclear absorption. Note that excitations of nucleons inside the core are  Pauli blocked. 
 
This Letter is organized as follows: In Section II we summarize the main characteristics of CENNS; in Section III we present our proposal to use the M\"ossbauer technique to observe CENNS and, in Section IV, we calculate the isomeric shift correction when a neutrino reactor flux are present. Section V summarizes our main conclusions.

\section{The Main Characteristic of CENNS}

\begin{figure}[H]
 \begin{center}
 \includegraphics[scale=0.5]{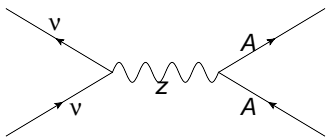}
 \caption{Feymann diagram of the  CENNS process.}
 \label{diagram}
 \end{center}
 \end{figure}

 The CENNS was proposed theoretically by Freedman~\cite{freedman}in 1974. A Feynman diagram of this weak process is shown in Fig.~\ref{diagram}. The effective Lagrangian to describe the process is given by 
\begin{equation}
L = G_{F} L^{\mu}J_{\mu},
\end{equation}
where $G_{F}$ is the Fermi constant, $L^{\mu}$ the lepton current, and $J_{\mu}$ is the hadron current inside the nucleus. 

Experimental efforts have been developed in the detection of CENNS, some of them represented by large scientific collaborations namely, COHERENT \cite{coherent,akimov}, CONNIE \cite{connie} and TEXONO \cite{tex}, among others. As mentioned before, after decades of searching, only in the last year the COHERENT Collaboration~\cite{akimov} announced the first irrefutable detection of CENNS.

The CENNS coherence, as it is well known, requires $qR \ll 1$, with $q$ being the transferred momentum and $R$ the nuclear radius. This implies that the wavelength of neutrinos will be comparable to the nuclear radius. Detailed discussions about the phenomena can be found in Refs.~\cite{freedman,mosel,krauss,formaggio,brice} and references therein. We stress the fact that the cross section of this process has the largest value $(\sigma \approx 10^{-38}$ cm$^{2})$ at least four orders of magnitude larger than other neutrino interactions in the same low-energy regime~\cite{formaggio}.

The Freedman differential cross section for this process is~\cite{freedman,giunt,kate}
\begin{equation}
\frac{d\sigma_{\rm CENNS}}{dT}=\frac{G_{F}^{2}}{4\pi}Q_{w}^{2}M_{A}F^{2}(q^{2})\bigg(1 - \frac{M_{A}T}{2E_{\nu}^{2}}\bigg) ,
\label{cross} 
\end{equation}
where $T$ is the transferred energy to the nucleus, ${A}$ is the target mass, $E_{\nu}$ is the neutrino energy and $Q_{w}= N -Z(1-4$sin$^{2}\theta_{w})$ is the weak charge, which depends on the number of neutrons ($N$) and protons ($Z$). Here $\theta_{w}$ is the Weinberg angle satisfying sin$^{2}\theta_{w}\approx 1/4$ and the proton contribution is negligible. The last fact made CENNS a very sensitive probe to nuclear neutron density~\cite{balant}. The form factor $F(q^{2})\to 1$ as $q\to 0$ define the coherence condition. 

We note that as a weak process, the interaction involved in the CENNS should be many order of magnitude greater than the gravitational phenomena. Even so, the MS was successfully choose as an appropriated technique to measure the gravitational red shift of light by Pound and Rebka \cite{pound,snider}, at the end of the fifties. Thus, we hopefully expected that MS can detect CENNS when properly employed.   

\section{The M\"ossbauer Technique Applied to Detect CENNS} 

One of the main characteristics of the MS is that the nuclei in the absorber material of the machine  are recoilless when interacting with photons that came from the source decay.  This condition is fundamental for the resonant radiation absorption in the MS. This feature was preserved for the CENNS because the energy is in the same range of the one of the photon. The transferred momentum by the $Z^{0}$ exchange is assumed to be transmitted to the valence neutron, slightly modifying the neutron distribution in the nuclear surface and promoting a typical isomeric shift correction in the MS experiment. In addition, it is straightforward to show that the recoilless nuclei in the source and absorber is consistent with the resonant radiation absorption and with coherent neutrino scattering by the nuclei. 

The fraction $f$ of the recoilless nuclei in $Z^{0}$ exchange between neutrino and nuclei in the CENNS can be analyzed similarly to the case of the  gamma radiation interaction. It can be shown that the fraction of recoilless events can be put in the form of Debye-Waller factor~\cite{marf}, which, for the CENNS, takes the form
\begin{equation}
f=\exp{\left(-\frac{T^{2}}{Mc^{2}\hbar\omega}\right)} ,
\end{equation}
where $T=E_{\nu}^{2}/2Mc^{2}$ is the energy transferred by the $Z^{0}$ to target nucleus of mass $M$. Here $\hbar\omega \approx 10^{-3} $ eV for Fe, Co etc, is the order of magnitude of energy lattice vibrations. In the range of neutrino energies below $\approx 50$~MeV the recoilless $f$ factor is essentially unity. Therefore, we argue that this small energy fraction is accommodated by the neutron levels.

\section{Isomeric Shift Correction due to the CENNS Interaction}

We consider that the energy transfer to the  valence neutron is done in a perturbative way, taking into account the first-order perturbation series in a shell model picture.
We assume that the perturbed neutron wave functions acquire a small projection in the next shell model state of the unperturbed system. This picture allows us to consider the problem as a two-level system, with the neutron state fluctuating between the two unperturbed state levels. In the present case we will focus on $^{57}$Fe because it is the most common in the literature, but many other nuclei can be studied with this technique, e.g., La, Te, Cd and Sm \cite{greenwood}.
The unperturbed $^{57}$Fe valence neutron is at a state of definite angular momentum, given by the common distribution of the neutron and proton content \cite{weiss} in the conventional nuclear shell model -- its wave function is regular at the origin (typically a Bessel function). Thus the perturbed valence neutron states, after the $Z^{0}$ interaction are 
\begin{eqnarray}
\Phi^{+}= \frac{-\lambda j_{3/2}(kr)}{\sqrt{1+\lambda^{2}}}+\frac{j_{5/2}(kr)}{\sqrt{1+\lambda^{2}}},\\ 
\Phi^{-}= \frac{j_{3/2}(kr)}{\sqrt{1+\lambda^{2}}}+\frac{\lambda j_{5/2}(kr)}{\sqrt{1+\lambda^{2}}}.
\label{pert}
\end{eqnarray}

The $\lambda$ parameter in the above equations appears in the perturbative treatment and is associated to the ratio between the energy transferred in the CENNS interaction~\cite{kuchiev}. Explicitly we have, 
\begin{equation}
\lambda = \lambda(E_{\nu})=\frac{3E_{\nu}^{2}}{8Mc^{2}(E_{5/2}-E_{3/2})}. 
\label{lambda}
\end{equation}
The term $(E_{5/2}-E_{3/2})$ is the difference between the energy of the non-perturbed states of the valence neutron. In $^{57}$Fe is this is 14.4~KeV.

In the context of the shell model for Woods-Saxon potential with spin orbit term~\cite{samuel,weiss,charm,kuksa}, the two states of valence neutron for the $^{57}$Fe  are represented by $j_{3/2}(kr)$ and $j_{5/2}(kr)$. We have used for the wave number of the valence neutrons $k \approx 0.5$ fm$^{-1}$.

The isomeric shift can be calculated~\cite{greenwood} as being
\begin{equation}
\delta I^{*}_{s}=\frac{4\pi Ze^{2}R^{2}}{5}\Bigl(\frac{R_{exc}-R_{gs}}{R}\Bigr)[\psi_{l=0}^{2}(0)_{a}-\psi_{l=0}^{2}(0)_{s}] ,
\label{deltas}
\end{equation}
where $Z$ is the number of protons in the nucleus, $R$ is the mean radius of the charge distribution and $R_{exc/gs}$ for the $Z^{0}$ excited and ground-state nuclear radius, respectively, and the $\psi 's$ is the $s$ electrons wave functions, evaluated at the origin~\cite{marf,greenwood}. In the literature, the difference $(R_{exc}-R_{gs})$ in conventional gamma absorption is reported to  be of order $10^{-3}R$ \cite{marf,greenwood}. Our estimative for $\frac{R_{exc}-R_{gs}}{R}$, calculated using the perturbed neutron wave functions $\Phi_{+/-}$ above is 
$\approx 10^{-4}R$. 
With this result and Eq. \ref{deltas}, we can obtain the correction at isomeric displacement induced by $Z^{0}$ and we see that it is only one order of magnitude smaller than the typical already measured values for the characteristic $\gamma$ measurements. This value for $\delta_{I^{*}_{s}}$ is perfectly solved with the MS technique accuracy, namely $10^{-10}$
eV \cite{pound,snider,greenwood}.  Consequently, we point out that if we take subtraction of a MS measurement without the neutrino flux and other result of identical measure with the reactor neutrino beam, we would reveal the contribution of the CENNS interactions.

\section{Conclusions}

In this work we develop some arguments that supports that MS could be a suitable technique to see the effect of the neutral interaction between neutrinos and nucleus. 

The correction obtained for this isomeric contribution at the MS experiment, e.g., $\sim 10^{-7}$~eV for $^{57}$Fe, appears to be greater than the energy resolution of this technique, which is $\sim 10^{-10}$~eV  , and we argue that in future this technique could be suitable to integrate the neutrino experimental search plants. 
\vspace{.91cm}
\begin{acknowledgments}
S. B. Duarte acknowledge financial support from CNPq. The authors are also grateful to Pedro Cavalcanti Malta, Gustavo Pazzini de Brito, José A. Helay\"el-Neto, H\'elio da Motta and Arthur M. K\'os for the encouragement and for clarifying discussions during the development of this work. 
\end{acknowledgments}

\end{document}